\documentclass[aps,prapplied,reprint,10pt,twocolumn,superscriptaddress]{revtex4-2}
\usepackage[utf8]{inputenc}
\usepackage{amsmath}
\usepackage{amssymb}
\usepackage{bbm}
\usepackage{graphicx}
\usepackage[countmax]{subfloat}
\usepackage{framed}
\usepackage{color}
\usepackage{hyperref}
\usepackage{epstopdf}
\usepackage{dsfont}
\usepackage{amsthm}
\usepackage{wrapfig}
\usepackage{relsize}
\usepackage{bm}
\usepackage{enumitem}
\usepackage{soul}
\usepackage{pifont}
\usepackage{todonotes}

\DeclareMathOperator{\Tr}{Tr}

\def\Tr{\hbox{Tr}}

\newcommand{\ket}[1]{\vert#1\rangle}

\newcommand{\ba}{\begin{eqnarray}}
\newcommand{\ea}{\end{eqnarray}}

\begin{document}

\title{\textit{Ab-initio} experimental violation of Bell inequalities} 

\author{Davide Poderini}
\author{Emanuele Polino}
\affiliation{Dipartimento di Fisica, Sapienza Universit\`{a} di Roma, Piazzale Aldo Moro 5, I-00185 Roma, Italy}
\author{Giovanni Rodari}

\affiliation{Dipartimento di Fisica, Sapienza Universit\`{a} di Roma, Piazzale Aldo Moro 5, I-00185 Roma, Italy}
\author{Alessia Suprano}

\affiliation{Dipartimento di Fisica, Sapienza Universit\`{a} di Roma, Piazzale Aldo Moro 5, I-00185 Roma, Italy}

\author{Rafael Chaves}
\affiliation{International Institute of Physics $\&$ School of Science and Technology, Federal University of Rio Grande do Norte, 59078-970, P. O. Box 1613, Natal, Brazil}

\author{Fabio Sciarrino } 
\email{fabio.sciarrino@uniroma1.it}
\affiliation{Dipartimento di Fisica, Sapienza Universit\`{a} di Roma, Piazzale Aldo Moro 5, I-00185 Roma, Italy}

\begin{abstract}
The violation of a Bell inequality is the paradigmatic example of device-independent quantum information: the nonclassicality of the data is certified without the knowledge of the functioning of devices. In practice, however, all Bell experiments rely on the precise understanding of the underlying physical mechanisms. Given that, it is natural to ask: Can one witness nonclassical behaviour in a truly black-box scenario?
Here we propose and implement, computationally and experimentally, a solution to this ab-initio task. It exploits a robust automated optimization approach based on the Stochastic Nelder-Mead algorithm. Treating preparation and measurement devices as black-boxes, and relying on the observed statistics only, our adaptive protocol approaches the optimal Bell inequality violation after a limited number of iterations for a variety photonic states, measurement responses and Bell scenarios. In particular, we exploit it for randomness certification from unknown states and measurements. Our results demonstrate the power of automated algorithms, opening a new venue for the experimental implementation of device-independent quantum technologies.
\end{abstract}

\maketitle
\section{Introduction}
The experimental guidance is quintessential in science. Not only empirical evidence is crucial to the development of models but conversely allows to test and improve theories. To interpret the data, however, one needs well understood and calibrated instruments. And for that, theoretical assumptions are always needed, creating a seemingly unstoppable circular argument. It is thus surprising that in the device-independent framework \cite{pironio2016focus}, quantum information tasks can be achieved simply from the data and without the need of precise knowledge of the devices.  

The paradigmatic example of device-independence is the violation of a Bell inequality \cite{bell1964einstein}. Not only it provides the most radical departure of quantum theory from classical concepts but also paves the way for applications ranging from cryptography \cite{acin2007device}, randomness certification \cite{acin2016certified}, self-testing \cite{STreview} and communication complexity \cite{buhrman2010nonlocality,ho2021quantum}. In principle, all of that is achieved simply by imposing the causal structure of the experiment but without any knowledge of the quantum states being prepared neither the measurements performed on them. In practice, however, all Bell experiments performed to date exploit the precise knowledge of the physical platform under use to maximize the Bell inequality violation (see for instance \cite{Shalm,Giustina,Hensen,christensen2015exploring}). Otherwise, how could one optimize the experimental setting in order to extract its most non-classical features? If our device is really treated as a black-box, how can we ever prove its quantum nature? 

Apart from its relevance in Bell scenarios and related topics, solving such a question might find application in range of fields such as quantum gravity \cite{marletto2017gravitationally,bose2017spin}, and quantum biology \cite{lambert2013quantum}, scenarios in which the detection of quantum effects can be a crucial task. But in all these cases, the quantum mechanisms at play, if any, might not be well understood. That is, simply by probing the system in different but not fully understood ways and looking at its response we should be able to witness its non-classical nature. 

Here we propose an adaptive automated optimization protocol exactly to solve this ab-initio task, in particular focusing on optimizing the violation of Bell inequalities without any prior knowledge on the quantum system and measurements. That is, in a fully black-box scenario. We exploit a Stochastic Nelder-Mead algorithm \cite{chang2012stochastic} that is an efficient, noise-resistant and  gradient-free algorithm for the search of global optima of functions. By repeatedly tuning the measurements parameters and  collecting statistics, the adaptive algorithm is able to optimize the Bell inequality violation after a limited number of measurements. Remarkably, the protocol is also robust against Poissonian fluctuations. We provide the proof of the applicability of our method by performing simulations and photonic experiments involving different bipartite states and measurement responses as well as a range of different Bell inequalities. After few hundreds of measurements the algorithm is able to approach the maximum possible violation for a given entangled state.

It is worthy pointing out that machine learning techniques are spreading out as powerful tools for quantum information tasks, in particular in the study of Bell nonlocality \cite{canabarro2019machine, krivachy2020neural, bharti2019teachg, wallnofer2020machine, dunjko2018machine}. For instance, reinforcement learning has been used for finding the optimal quantum violations of  Bell inequalities for many-body systems with fixed known measurement settings \cite{deng2018machine} and  for the design of optical experiments optimizing Bell inequality violations~\cite{melnikov2020setting}. In such cases, however, a precise quantum description was required. To our knowledge, our method is a novel way to detect and optimize Bell nonlocality in a truly black-box situation.
To showcase its applicability, we exploit it for maximizing the certified randomness extraction from unknown system and measurements, thus opening a fruitful venue for the device-independent quantum information framework.

\begin{figure}[t!]
    \centering
    \includegraphics[width=.9\columnwidth]{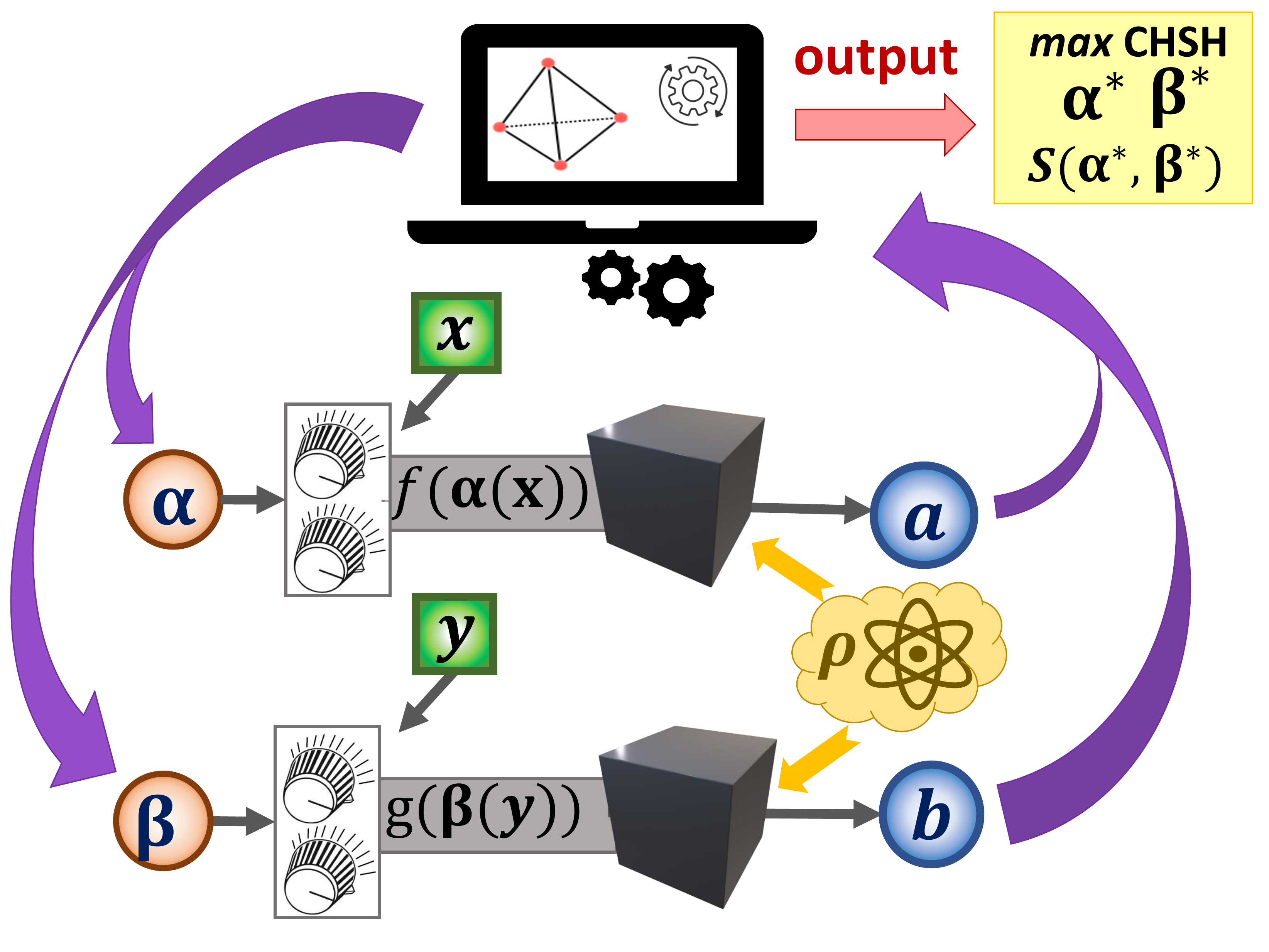}
    \caption{\textbf{Bell inequality optimization through Stochastic Nelder-Mead algorithm.} 
    An initial simplex in the parameters space  is chosen. At each iteration the algorithm performs tests of the Bell inequality $S \leq L$ using measurement parameters chosen accordingly to previous results. At the end of the protocol we reach values close to the optimal measurements parameters $\bm{\alpha}^{*}$ and $\bm{\beta}^{*}$ maximizing the Bell inequality. Measurements and states are treated as black-boxes, thus one has to choose functions $f(\cdot)$ and $g(\cdot)$ mapping the parameters to the empirically observed value of $S$.}
    \label{fig:nelder}
\end{figure}

\section{Violation of Bell inequalities as a black-box optimization problem} Virtually any experiment in quantum physics can be understood as an instance of a prepare and measure scenario. Physical systems described by a quantum state are prepared and measurements are used to reveal its statistical properties. Depending on the application, different levels of control and characterization over the preparation and measurement devices can be allowed. In quantum tomography \cite{d2003quantum}, for instance, the unknown quantum state being prepared can be reconstructed if we trust and know our measurement apparatus. In other cases \cite{buscemi2012all}, the preparation can be assumed to be known while the measurements cannot. In the context of quantum information, the more we assume the more open is the way to malicious attacks \cite{lydersen2010hacking} or wrong conclusions \cite{rosset2012imperfect}.

In this sense, the violation of a Bell inequality  provides the ultimate security level, since by assuming only the causal structure imposed to the experiment, but no knowledge of the preparation and measurement devices, one can infer a number of properties of the physical system under test \cite{Brunner}. More precisely, in a Bell test a number of distant parties receive shares of a quantum system prepared by an uncharacterized device. Upon receiving their share, they can locally manipulate and measure them using again unknown devices. That is, each of them have a black-box with knobs that can be controlled, but the effect of these knobs within each device is unknown.

Consider the bipartite case involving the observers Alice and Bob. At each run of their experiment, input variables $x$ and $y$ decide how the knobs will be turned, leading to the corresponding measurement outcomes $a$ and $b$, respectively (see Fig. \ref{fig:nelder}). The experimental data is thus encoded into the conditional probability distribution $p(a,b \vert x,y)$.

\begin{figure}[t!]
\includegraphics[width=0.99\columnwidth]{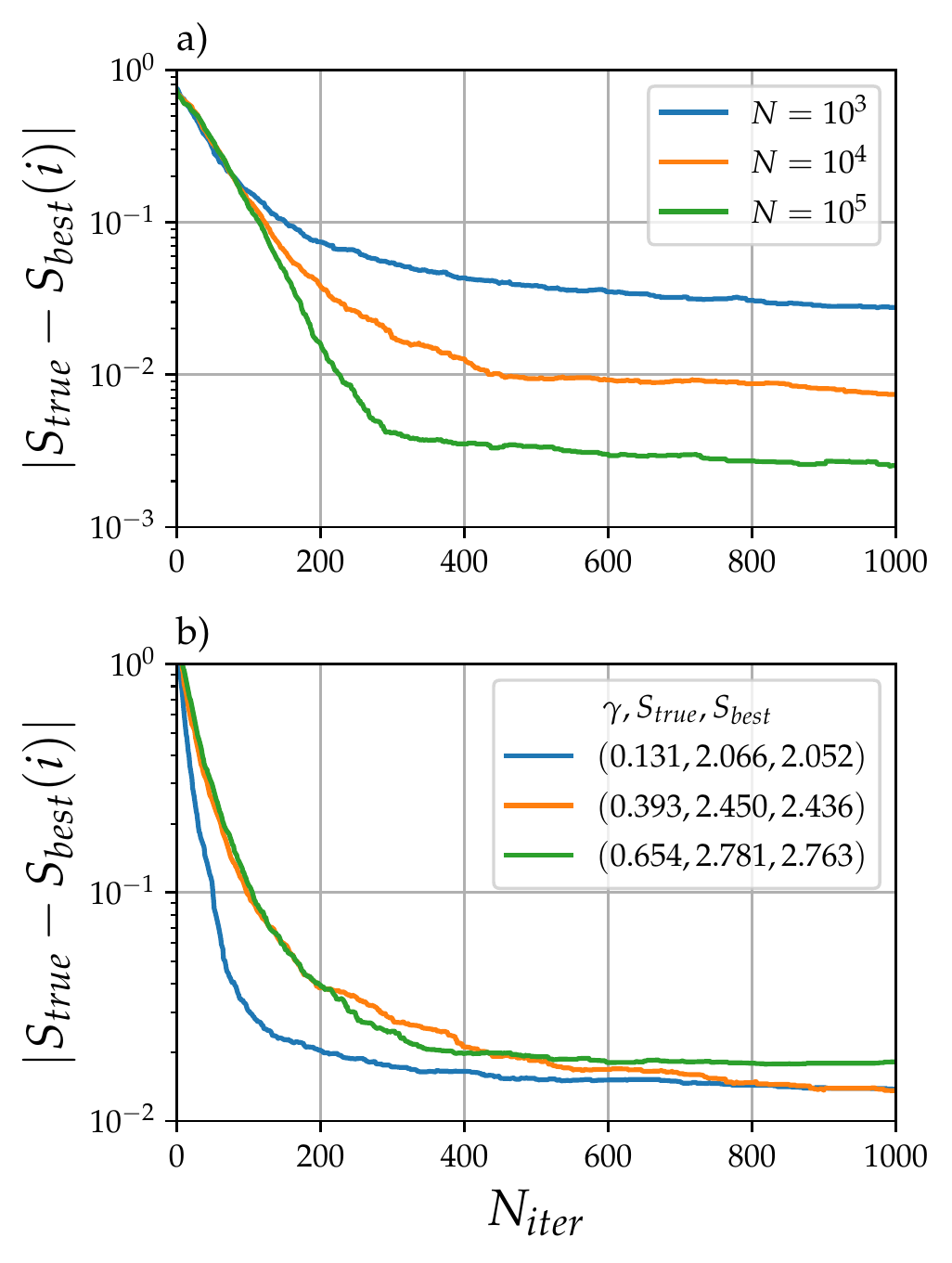}
\caption{\textbf{Simulation Results.} 
The difference $|S_{true}-S_{best}|$ is plotted as function of the algorithm's iteration. For all plots the curves are obtained by averaging over 100 simulation runs.
\textbf{a)} Singlet state with different numbers of events $N$. \textbf{b)}  Pure unbalanced 
states (parametrized by $\gamma$ in radians). The label $S_{true}$ specifies the optimal violation of the CHSH inequality for the corresponding states. The mean number of Poissonian events relative to each measurement is $10^5$.  Our approach 
works also for states with violation as low as $S_{true}=2.066$ achieving $S_{best}=2.052$, after $1000$ iterations.
}
\label{fig:sim}
\end{figure}

In a quantum description of such experiment,  their observations are described as $p(a,b\vert x,y)= \mathrm{Tr} \left[(M_a^x \otimes M_b^y)\rho  \right]$, where $\rho$ describes the quantum state being prepared and $M_a^x$ and $M_b^y$ the measurement operators. Classically, however, the assumption of local realism \cite{Brunner} implies that the experimental data can be decomposed as
\begin{equation}
\label{eq:lhv}
p(a,b \vert x,y)= \sum_{\lambda} p(\lambda) \; p(a\vert x,\lambda) \; p(b\vert y,\lambda) \;.
\end{equation}

Bell's theorem \cite{bell1964einstein} implies that some quantum predictions are incompatible with the classical prescription \eqref{eq:lhv}. This is the phenomenon known as quantum non-locality that can be witnessed by the violation of a Bell inequality, generally written as a linear constraint over the observed probabilities given by
\begin{equation}
\label{eq:Bellineq}
S=\sum_{a,\, b,\,x,\, y}  \alpha_{a,\, b,\,x,\, y} \; p(a,b\vert x,y)\le L \; ,
\end{equation}
where $\alpha_{a,\, b,\,x,\, y}$ are integer coefficients and $L$ is the bound arising from the classical description \eqref{eq:lhv}.

Checking if $p(a,b \vert x,y)$ violates a Bell inequality we can conclude, for example, that the shared state has to be entangled and that measurements being performed are incompatible \cite{STreview}. That is, we can probe some properties of the system under test in a device-independent manner. In practice, however, the quantum state $\rho$ and the measurement operators $M_a^x$ and $M_b^y$  have to be tuned precisely in order to obtain and maximize a Bell inequality violation. And, in a number of situations, such knowledge contradicts the basic black-box assumption of the scenario. In what follows we will propose a solution to deal with that.

The inputs $x$ and $y$ tell the parties how to turn their knobs in their measurement devices. For instance, in a photonic implementation using the polarization degree of freedom of photons, the measurements are performed by changing the orientation of half and quarter wave-plates. Similarly, the degree of entanglement of the source that we exploit can be tuned by changing the pump polarization. In a black-box approach, each of these changes in the preparation and measurements are simply described by a knob that can be turned. But what this turning of knobs is doing inside the devices we cannot speak of. The knobs for Alice and Bob are described by a set of (continuous or discrete) variables $\bm{\alpha}$ and $\bm{\beta}$, each a function of the inputs $x$ and $y$, respectively. In turn, the source is described by $\bm{\gamma}$. That is, $M_a^x=M_a(\bm{\alpha(x)})$, $M_b^y=M_b(\bm{\beta(y)})$ and $\rho=\rho(\bm{\gamma})$.

\begin{figure*}[ht!]
\includegraphics[width=0.9\textwidth]{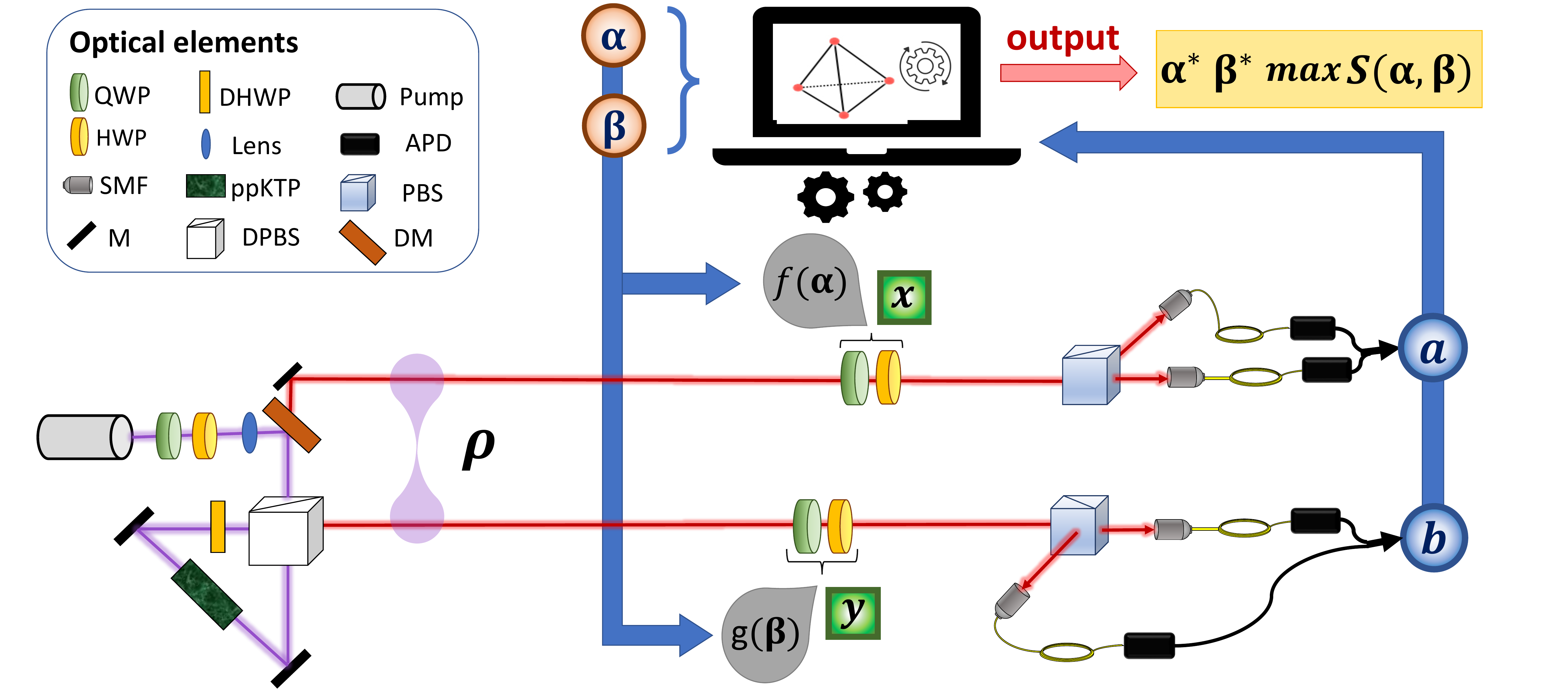}
\caption{\textbf{Experimental setup} 
A Sagnac-based source generates photonic pair states of the form $\cos \gamma \ket{HV}+e^{i \phi} \sin  \gamma \ket{VH}$ unknown to the optimization algorithm. 
The polarization of the two photons is measured by two
waveplates whose angles are determined by unknown functions, $f(\cdot)$ and $g(\cdot)$ of known parameters $\bm{\alpha}$ and $\bm{\beta}$. 
New parameters $\bm{\alpha}$ and $\bm{\beta}$ are chosen based on the results $a$ and $b$ of the previous iteration. At the end of the process, optimal values $\alpha^*$ and $\beta^*$ are chosen in order to optimize the Bell inequality violation. QWP: quarter-waveplate; HWP: half-waveplate; DHWP: dual wavelength half-waveplate; PBS: polarizing beamsplitter; DPBS: dual-wavelength polarizing beam splitter; SMF: single mode fiber; APD: avalanche photodiode single photon detector; ppKTP: periodically poled potassium titanyl phosphate; M: Mirror; DM: Dichroic Mirror.}
\label{fig:expsetup}
\end{figure*}

Given a Bell inequality described by $S=S(\bm{\alpha},\bm{\beta},\bm{\gamma})$, our objective will be to maximize $S$, particularly obtaining a value that surpasses the local bound $L$. Importantly, we want to achieve that without knowing how function $S$ depends on the knob parameters $\bm{\alpha}$, $\bm{\beta}$ and $\bm{\gamma}$.
That is, the optimization is performed in a truly black-box scenario. As we will show, changing iteratively the value of the parameters and observing, in real time, the change of $S$ we will be able to reach its optimal value after a remarkably low number of iterations.

We employ a gradient-free and direct search algorithm, the Nelder-Mead simplex method \cite{nelder1965simplex} and its stochastic variant \cite{chang2012stochastic},  able to efficiently optimize a multidimensional function by simply comparing some of its values. This is achieved even in the presence of noise, unavoidable in an experimental implementation. In the following we will fix the parameter $\gamma$ describing the source but our method can also be used to optimize over it.

The algorithm adaptively evolves a simplex \cite{note1} living in the space of the parameters $\bm{t}=( \bm{\alpha},\bm{\beta}) \in \mathcal{T}$.  The Bell function $S(\bm{t})$ can be calculated for each of the $n$ simplex points: $\{\bm{t}_1,...,\bm{t}_n \} \subseteq \mathcal{T}  \longrightarrow \{S(\bm{t}_1),...,S(\bm{t}_n) \}$.
As input of the optimization protocol an initial simplex $\Omega_0=\{\bm{t}^0_1,...,\bm{t}^0_n\}$ is generated through \emph{Latin Hypercube Sampling} \cite{chang2012stochastic}.
Starting from the initial simplex, the algorithm repeats an optimization cycle that updates the simplex (Fig \ref{fig:nelder}). Each cycle of the adaptive algorithm is composed of three main steps: 1) Sorting the points of the simplex $\Omega_{k-1}$, based on the values of the associated cost function such that the worse point is deleted, 2) from the barycenter of the simplex, a new point $\bm{t}^{ref}$ is generated through a reflection rule, 3) according to the cost of $\bm{t}^{ref}$, the simplex is updated by including $\bm{t}^{ref}$ itself or geometrically generating a further point. If such point is not promising enough an \textit{Adaptive Random Search} is employed to generate a point $\bm{t}^{ARS}$ \cite{note2}. A more detailed account of the above steps is provided in the Appendix A, while a comparison between the the devised algorithm and standard non-adaptive approaches is provided in Appendix B.

\section{Simulation and numerical tests on the CHSH inequality}
To illustrate the main features of our approach , in the following we will focus on the Clauser, Horne, Shimony and Holt (CHSH) scenario \cite{clauser1969proposed} consisting of dichotomic inputs and outputs. The only class of Bell inequalities in this scenario is given by
\begin{equation} \label{chsh}
\begin{split}
S=\langle  A_0 B_0\rangle+\langle  A_0 B_{1}\rangle+\langle  A_{1} B_0\rangle-\langle  A_{1} B_{1}\rangle \le2\;,
\end{split}
\end{equation}
where $ \langle  A_x B_y\rangle \equiv  \sum_{a,b}\,  a\,b\,p(a,b|x,y)$ is the expectation value (given the inputs $x$ and $y$) of Alice and Bob outcomes  $(a,b) \in \{+1, -1\}$. Moreover our approach can be applied for general optimization and we consider a range of different scenarios such as the chained \cite{braunstein1990wringing}, tilted \cite{acin2012randomness} and the non-linear Tsirelson-Landau-Masanes inequalities \cite{masanes2003necessary}.

We performed simulated optimizations and experiments covering a range of pure and noisy quantum states. Here we focus on quantum states given by
\begin{equation}
    \ket{\psi^\gamma} = \cos \gamma \ket{HV} + \sin \gamma \ket{VH}
    \label{eq:stategen}
\end{equation}
In turn, the most general projective measurement on a qubit state is given by $
\hat{P}(\theta,\phi) = \vec{r}(\theta,\phi) \cdot \vec{\sigma}$, where $\vec{r}(\theta,\phi)$ represents a unit vector in polar coordinates and $\vec{\sigma} = (\sigma_x,\sigma_y,\sigma_z)$ are Pauli operators. Considering that the input of the CHSH test defines the angles $\theta, \phi$ of such measurements (noticing that this knowledge is never used by the algorithm), we have at least a total number of $8$ measurement parameters ($2$ for each measurement of Alice and Bob). This defines the parameter space where the algorithm searches for the CHSH optimization.

\begin{figure}
\includegraphics[width=.9\columnwidth]{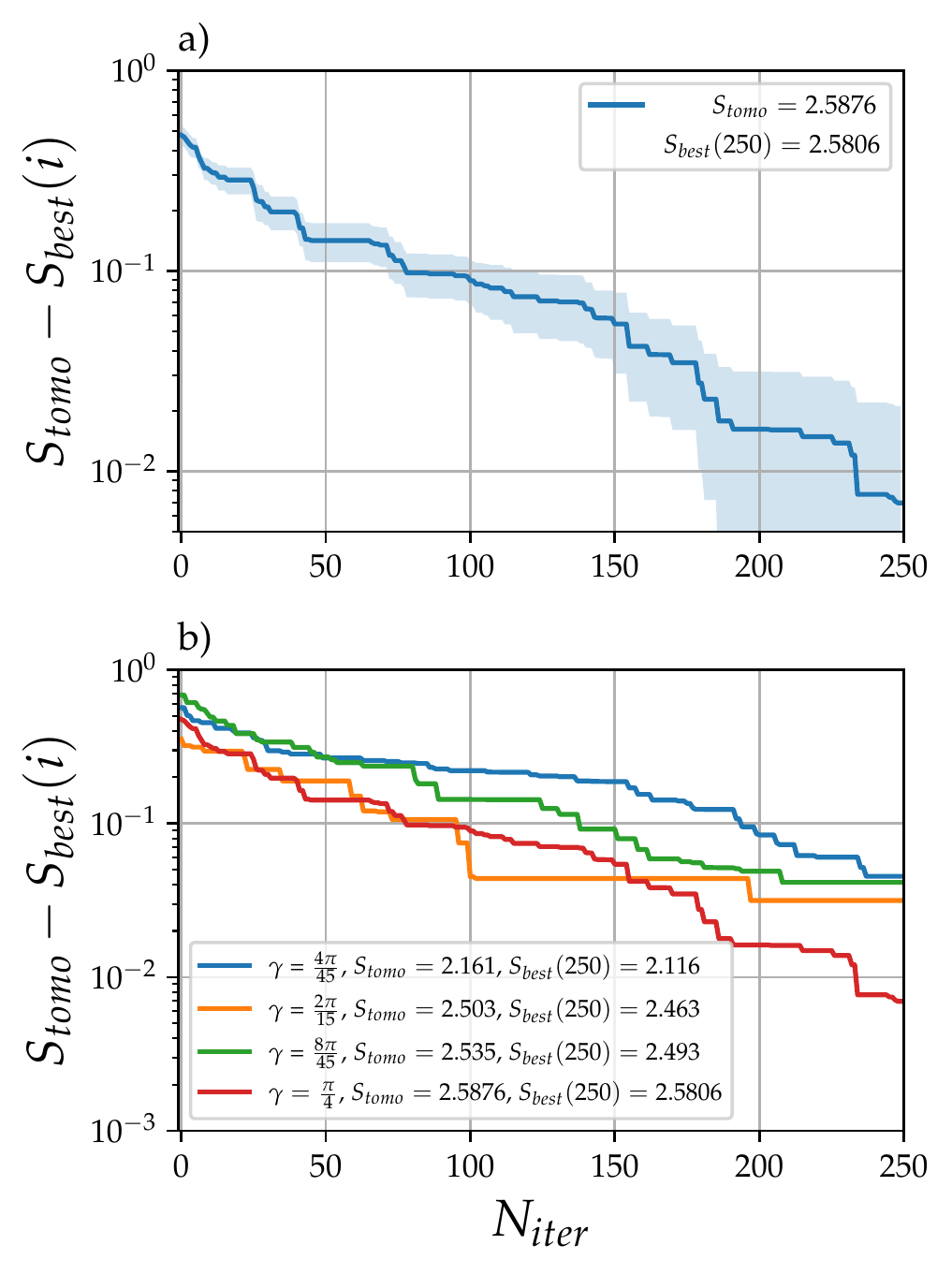}
\caption{\textbf{ Experimental Results} Average difference $S_{tomo}-S_{best}$ between the value $S_{tomo}$ that is the maximum CHSH violation achievable by the state from the quantum state tomography and $S_{best}$ the best value obtained by the algorithm  as function of the iteration of the algorithm.
\textbf{a)}  The shaded blue area represents the standard deviation of the mean. The mean number of events for each CHSH measurement is $\sim 4 \cdot 10^4$.  $S_{tomo} = 2.588 \pm 0.006$, $S_{best} =  2.581 \pm 0.014$. 
\textbf{b)} Different values of $\gamma$ parameter relative to the state are chosen:
$ 4/45\pi, 2/15 \pi, 8/45 \pi, \pi/4$.
For all the data the number of parameters is 8 (4 for each station) and the mean number of collected photons for each CHSH measurement is $\sim 4 \cdot 10^4$. Each curve is the average of 5 repetitions of the optimization (see Table \ref{tab:tablepar}). 
}
\label{expres}
\end{figure}

It is worth pointing out that, since we are in a fully black-box scenario, the optimization process does not require the exact form of the quantum state or measurements, that are only chosen for simulation purposes. As a matter of fact, in the optimization, one has to choose how the algorithm will map the parameters $\bm{\alpha}$ and $\bm{\beta}$ to the observable measured at a particular iteration. As discussed in Appendix C, we have tested different nonlinear response functions, such as the hyperbolic sine or a logistic function. Finally, To mimic the experimental situation, Poissonian fluctuations are added in the measurements of operators. Such fluctuations are characterized by the parameter $N$, corresponding to the total number of Poissonian events  used to calculate the CHSH parameter at each iteration.

The figure of merit of the optimization process is given by
$\Delta = |S_{true} - S_{best}(i)|$ , where $S_{best}(i)$ corresponds to the best value of the CHSH parameter inside the simplex at the $i$-th iteration step, while $S_{true}$ is the maximum CHSH value achievable by the considered state and that can be computed resorting to the Horodecki criterion \cite{horodecki1995violating}.

In figure \ref{fig:sim}a) one can see that increasing $N$ (thus reducing the Poissonian fluctuations) not only leads to fast convergence but also to very small errors.
As shown in Fig. \ref{fig:sim}b) (considering $N=10^4$), for quantum states with different levels of entanglement, already with $200$ iterations the error reaches  values close to $\Delta=10^{-2}$.

\begin{table}
\begin{center}
\begin{tabular}{cccc}
 $\gamma$ & $S_{tomo}$ & $S_{best}$ & $\sigma$ \\
 \hline
 $\frac{4\pi}{45}\;\;$ & $2.161 \pm 0.026\;\;$ & $2.116 \pm 0.025\;\;$ & $0.056 \pm 0.020$\\ 
 $\frac{2\pi}{15}\;\;$ & $2.503 \pm 0.008\;\;$ & $2.463 \pm 0.024\;\;$ & $0.054\pm0.019$\\  
 $\frac{8\pi}{45}\;\;$ & $2.535 \pm 0.012\;\;$ & $2.493 \pm 0.031\;\;$ & $0.069 \pm 0.024$\\  
 $\frac{\pi}{4}\;\;$ & $ 2.588 \pm 0.006 \;\; $ & $2.581 \pm 0.014\;\;$ & $0.044 \pm 0.010$\\   
\end{tabular}
\end{center}
\caption{
\textbf{Experimental results:} The average Bell violations ($S_{best}$) for different states, identified by $\gamma$ as in \eqref{eq:stategen}, are listed alongside the reference violation found by tomographic reconstruction of the state ($S_{tomo}$).
Column $\sigma$ represents the standard deviation associated with distribution of the single optimization runs.
}
\label{tab:tablepar}
\end{table}

\section{Experimental ab-initio optimization tests} To demonstrate our protocol we exploited polarization states of pairs of photons.
 Entangled states of the form \eqref{eq:stategen} were generated through a type-II spontaneous parametric down conversion process inside a periodically poled  KTP  crystal (ppKTP), pumped by a continuous wave $\lambda=404$ nm laser in a Sagnac interferometric geometry \cite{kim2006phase,fedrizzi2007wavelength}. 

The photons exiting the source are measured through two polarization analysis stages, each composed of a quarter-waveplate (QWP) followed by a half-waveplate (HWP) and a polarizing beam splitter (PBS). The measurement parameters that are tuned determine the rotation angles of the waveplates by means of functions $f(\bm{\alpha})$ for Alice and $g(\bm{\beta})$ for Bob station (Fig. \ref{fig:expsetup}).

We  performed experimental  optimization tests varying the unknown states, number of parameters and measurement responses. In the experimental case we compare the highest reached violations $S_{best}(i)$ at $i$-th iteration with the maximum violation $S_{tomo}$ obtained by applying the Horodecki criterion \cite{horodecki1995violating} to the density matrix reconstructed with quantum state tomography \cite{james2005measurement}. Considering the maximally entangled state, Fig. \ref{expres}a show that we reach $\Delta < 10^{-2}$ after only 250 iterations, even considering an 8-parameter optimization. Experiments with different initial entangled states (Eq. \eqref{eq:stategen}) show a similar fast convergence (see Fig.\ref{expres}b). Remarkably, we achieve experimental Bell violations even for states with little entanglement and optimal CHSH violation as low as $S_{tomo}=2.161 \pm 0.026$ (see Table \ref{tab:tablepar} for the details).

\section{Beyond the CHSH inequality}
In the previous sections we have focused on the CHSH inequality. In what follows, we will show how our approach can be applied to a variety of different scenarios.

\begin{figure*}[ht!]
    \makebox[\textwidth][c]{\includegraphics[width=1.1\textwidth]{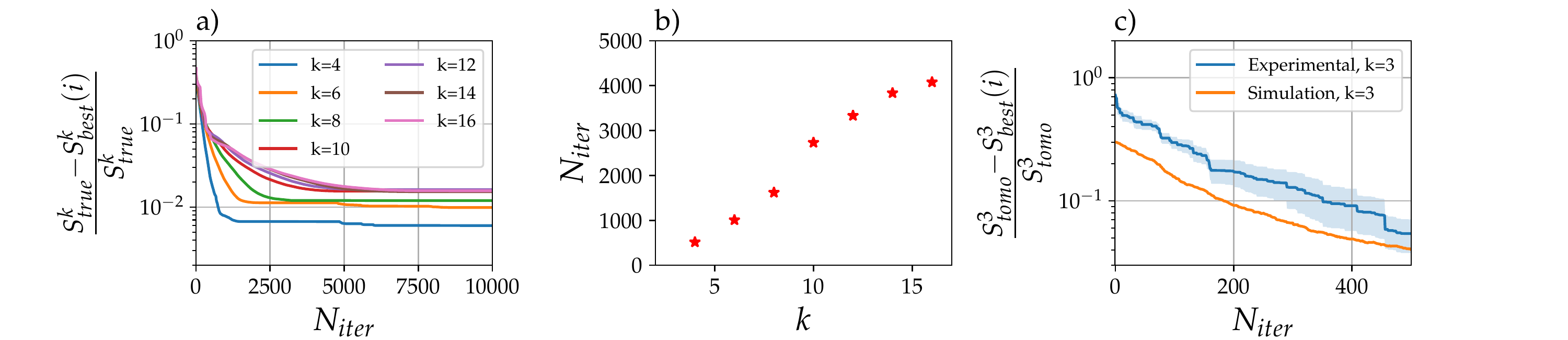}}
\caption{
\textbf{Chained Bell inequalities optimization (simulation and experiment).}
\textbf{a)} Mean results over 50 optimization runs of the chained Bell inequalities, computed considering a singlet state, increasing $k$ measurement per party and Poissonian measurement uncertainty with $N=10^6$. Since the achievable maximal violation $S^k_{true}$ varies with $k$, we report the rate $\frac{S^k_{true} - S^k_{best}}{S^k_{true}}$ as a function of the algorithm's iteration. The number of optimization parameters increases linearly and is given by $4k$. In all cases the algorithm is able to approach the expected minimum within $\approx 1\%$ after $\approx 5000$ noisy evaluations.
\textbf{b)} Median (over 50 repetitions) number of measurements needed  to reach a distance lower than $2\%$ between the ab-initio evaluation of the chained parameter $S^k$  and its maximum reachable value, as function of the number of settings $k$. As can be seen the number of iterations grows linearly with the number of optimization parameters showing the efficiency of our method.
\textbf{c)} Experimental optimization of the chain inequality with $k=3$, hence $12$ control parameters are necessary. We report the rate $\frac{S^3_{tomo} - S^3_{best}}{S^3_{tomo}}$, averaged over 5 experimental runs, as a function of the algorithm iteration.  The number of events for each measurement is $\sim 3 \cdot 10^4$. Here $S^3_{tomo}$ is again the maximum violation of $S^3$ achievable by the state from the quantum state tomography, computed through numerical optimization. The blue line indicates the average simulation (repeated 100 times) for  a Werner state with $p=0.90$, employing  $N=3\cdot 10^4$ Poissonian events with added Gaussian noise ($\sigma=0.05$)  to simulate the real experimental conditions, where 
    $S^3_{tomo} = 2.399 \pm 0.003$ and
for the average over the 5 runs we have $S^3_{best} = 2.28 \pm 0.02$ with a variance of $0.002$.
}
\label{fig:chain}
\end{figure*}

\subsection{Chained Bell inequalities}

\begin{figure}[ht!]
    \centering
    \includegraphics[width=\columnwidth]{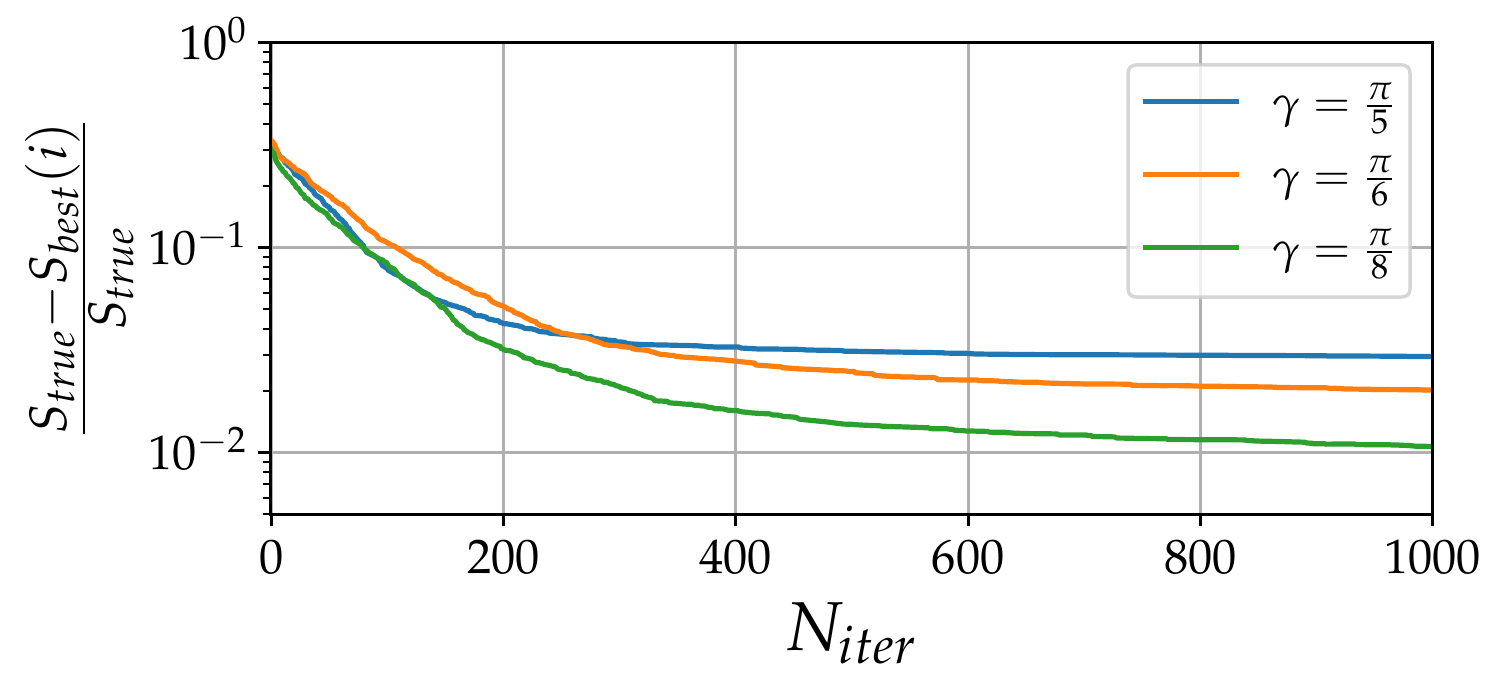}
    \caption{\textbf{Tilted Bell inequality black-box optimization.} Computational simulation considering a mean (over 100 repetitions) of the tilted Bell inequality in Eq. \eqref{eq:tilted} with parameters $\alpha=1$ and $\beta=2 - 2 \sin^2 \gamma$ optimized on the source state, simulating a Poissonian statistic with $N=10^4$ events for each measurement.
    }
    \label{fig:tilted}
\end{figure}

\begin{figure*}[ht!]
    \includegraphics[width=0.99\textwidth]{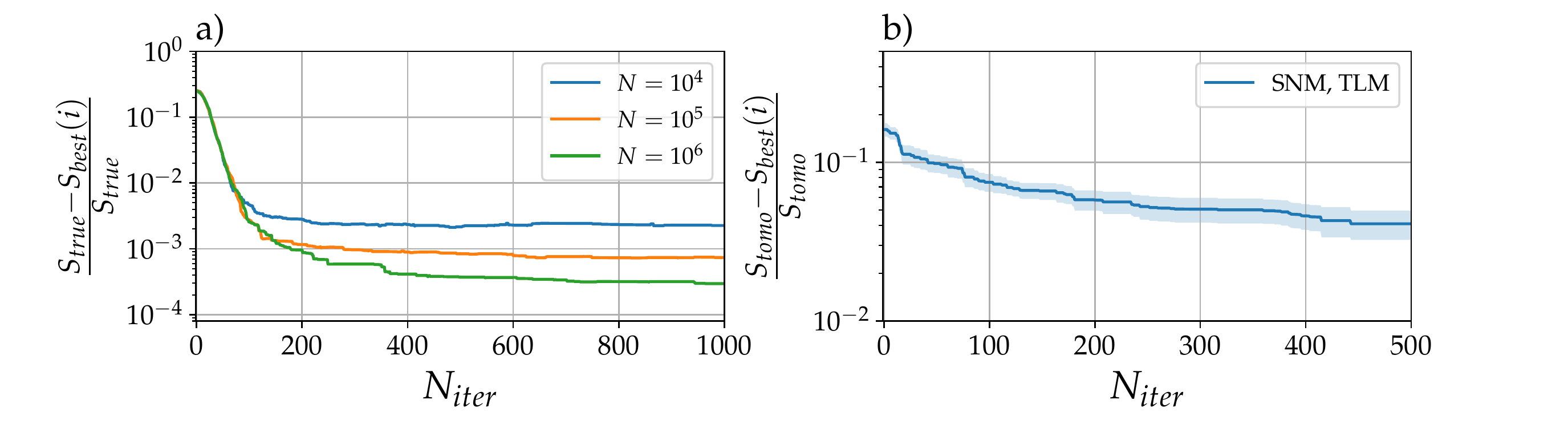}
\caption{\textbf{Optimization of quantum Bell inequalities (simulation and experiment).} 
\textbf{a)} The rate $\frac{S_{true} - S_{best}}{S_{true}}$ is plotted as a function of the algorithm's iteration. We report the mean results obtained by averaging 50 simulation runs of the SNM optimization of the TLM inequality \cite{masanes2003necessary} considering a pure singlet state. The different curves represent the distinct trends for three separate values of the Poissonian measurement uncertainty ($N$).
\textbf{b)} Experimental results of the optimization of the TLM inequality proposed in \cite{masanes2003necessary}. The difference $\frac{S_{tomo} - S_{best}}{S_{tomo}}$ has been averaged over $8$ experimental repetitions with 8 measurement parameters and it is reported as a function of the algorithms' iteration. The average of the optimal value found is $S_{best} = 2.862 \pm 0.007$, with a variance of $0.00039$, which is comparable with the one found by performing the full quantum state tomography $S_{tomo} = 2.98 \pm 0.01$. The average number of events for each measurement is $\approx 4.4 \cdot 10^4$. 
}
\label{fig:masanes}
\end{figure*}

An important class of bipartite Bell inequalities are the so-called chained Bell chained inequalities \cite{braunstein1990wringing}, given by
\begin{equation} \label{eq:classical_bounddd}
\begin{aligned}
    S^k &= \sum_{i=1}^{k}|I_{i}| \leq k-1\\
    \text{where} \; I_{i} &= \dfrac{1}{2} \sum_{x=i-1}^{i} \langle A^{x}B^{i-1} \rangle\;.
\end{aligned}
\end{equation}
with $A^{k}=-A^{0}$. 
Differently from the CHSH case, Alice and Bob can perform an arbitrary number of $k$ measurements each. This class of inequalities have found many applications, ranging from randomness amplification \cite{PhysRevA.90.032322} to self-testing of quantum states \cite{upi2016}.

Our simulation and experimental results for the chain inequality are shown in Fig. \ref{fig:chain}. Numerical simulations (Fig. \ref{fig:chain}a) show that even for increasing $k$ (up to $k=12$) the black-box optimization algorithm is still able to reach the values very close the maximum possible Bell inequality violation. Fig. \ref{fig:chain}b) shows that the number of iterations required to obtain a value lower than $2\%$ for the distance between the maximum reachable violation and the violation achieved by the optimization scales approximately linearly with $k$. Thus, our approach remains very efficient even by increasing the number of measurements and complexity of the Bell scenario. Finally, Fig. \ref{fig:chain}-c) shows our experimental results for $k=3$ (involving $12$ measurement parameters in total) also comparing it with a numerical simulation.

\subsection{Tilted Bell inequality}

Even though in the CHSH scenario the CHSH inequality is the only tight Bell inequality (a facet of the local set), it is known that non-tight Bell inequalities might play an important role for quantum information processing. For instance, while the maximal violation of the CHSH inequality allows tEven though in the CHSH scenario the CHSH inequality is the only tight Bell inequality (a facet of the local set), it is known that non-tight Bell inequalities might play an important role for quantum information processing. For instance, while the maximal violation of the CHSH inequality allows the generation of 1.23 bits of certified randomness

he generation of 1.23 bits of certified randomness
\cite{note4}, the use of different inequalities allows to get arbitrarily close to the maximum of 2 bits of certified randomness reachable by the CHSH scenario (two inputs and two outputs for each party). For that aim, the so called  tilted Bell inequalities \cite{acin2012randomness} have been introduced, allowing for the  optimal randomness certification in a Bell scenario even using non-maximally entangled states of the form $\cos\gamma \ket{00} + \sin \gamma\ket{11}$.
The general form of the inequality for local classical models, is given by
\begin{equation}
\begin{split}
    I^{\alpha}_{\beta} &= \beta \langle A_1 \rangle + \alpha \langle A_1 \, B_1\rangle +  \alpha \langle A_1 \, B_2\rangle + \langle A_2 \, B_1\rangle -\langle A_2 \, B_2\rangle \\
   &  \le 2 \alpha + \beta  \quad ,
\end{split}
\label{eq:tilted}
\end{equation}
 where $\alpha  \ge 1$ and $\beta \ge 0$ .
 
To gauge our approach in the violation of this inequality, we consider the case where $\alpha=1$ and $\beta=2 - 2 \sin^2 \gamma$ are optimized to maximize the violation of Eq. \eqref{eq:tilted}, whose maximum quantum value is $I^{\alpha}_{\beta}=2 \sqrt{(1+\alpha^2)(1+\beta^2/4)}$. 
As displayed in Fig. \ref{fig:tilted}, computational simulations considering different quantum states (parametrized by $\gamma$) show what our black-box optimization is able to approach the optimal value even with only $10^4$ Poissonian events per measurement. This shows that our black-box optimization might find applications in device-independent randomness certification as well. It is worthy pointing out that in this simulation we are already choosing the optimal inequality for the given state. In the future, one might consider the case where the underlying state is unknown and thus not only the measurements have to be optimized in a black-box scenario but also the objective function (the Bell inequality) can be optimized over.

\subsection{Black-box optimization of quantum inequalities }
 
 So far, we have considered only the case of classical Bell inequalities.
 In some cases, however, one might also be interested in exploring the boundary of the set of quantum correlations \cite{PhysRevA.97.022104,PhysRevA.99.032106,PhysRevX.5.041052}.
 
To illustrate the applicability of our approach also in this case, we consider the black-box optimization of the so called Tsirelson-Landau-Masanes (TLM) inequality \cite{masanes2003necessary} to bound the set of quantum correlations in the CHSH scenario. We consider a specific symmetry of these quantum inequalities given by
\begin{equation}
        - \pi \leq -\sin^{-1} x_1 + \sin^{-1} x_2 +\sin^{-1} x_3+ \sin^{-1} x_4 \leq \pi\\
        \label{eq:tlm}
\end{equation}

where $\bm{x}=(\langle A_0 B_0\rangle,\langle A_0 B_1\rangle,\langle A_1 B_0\rangle,\langle A_1 B_1\rangle)$. 

We simulated the optimization of this quantum inequality considering a singlet state and different numbers of Poissonian events (Fig. \ref{fig:masanes}-a). The goal here is to reach the optimal value of the function of correlations in Eq. \eqref{eq:tlm}, possibly getting as close as possible to the quantum bounds, which in the case of maximally entangled states is equal to $\pi$. Moreover, we have also performed an experimental test to demonstrate the performance of our method also in a real noisy setup (Fig. \ref{fig:masanes}-b). 
The black box optimization of the quantum inequality follows a similar trend, approaching its optimum value after a reasonably small number of iterations. Thus our approach can also find relevant applications in testing the limits of quantum predictions for possible deviations from quantum theory.

\section{Ab-Initio Randomness certification}
\begin{figure}[ht!]
    \includegraphics[width=.9\columnwidth]{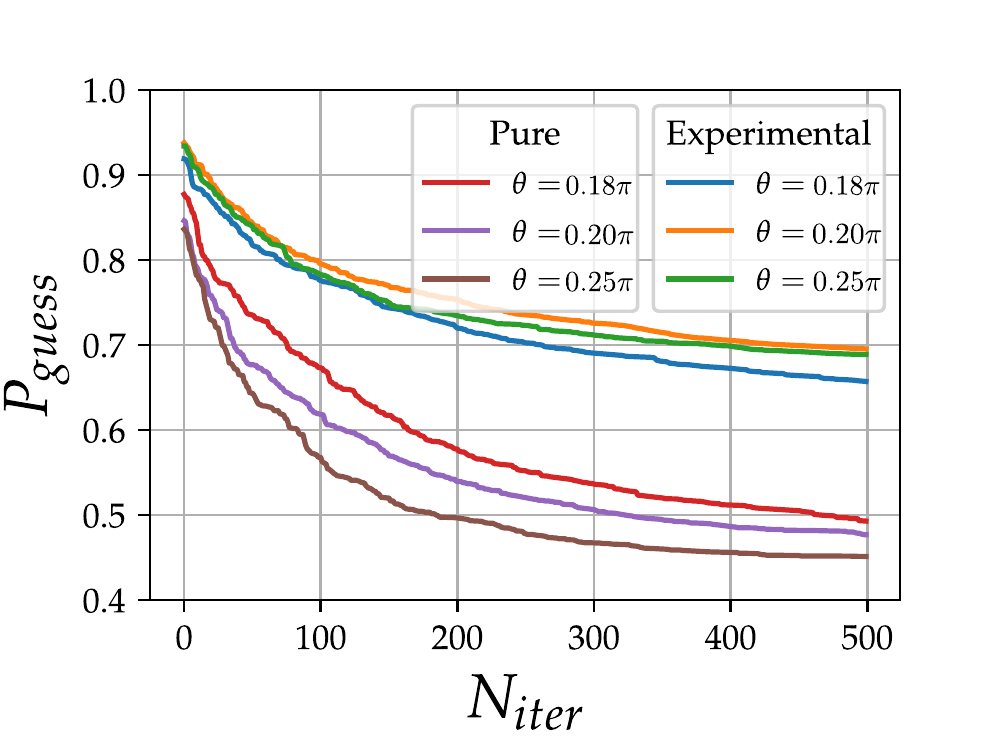}
\caption{\textbf{Computational simulation of extracted randomness} Optimal $p_{guess}$ of an adversary as a function of iterations for selected values of the parameter $\gamma$ for states of the form $\ket{\psi^\gamma} = \cos\gamma \ket{HV} + \sin\gamma \ket{VH}$. 
The experimental curves are obtained by considering density matrices of states reconstructed using experimental data. 
As in the main text, the behaviours used to evaluate Eve's $p_{guess}$ are obtained directly through Born's rule $p(a,b|x,y) = \Tr(A_{a|x}\otimes B_{b|y} \rho)$, where $\rho$ is the reconstructed density matrix.}
\label{fig:SIrandomness}
\end{figure}

Now, we show how our approach can be exploited to certify and maximize randomness, the paradigmatic application of Bell nonlocality \cite{acin2016certified,xu2020secure,Miller2016,Vazirani2012,ArnonFriedman2018}. 
While randomness can be certified via the violation of Bell's inequalities \cite{acin2016certified,herrero2017quantum,pironio2010random,colbeck2011private,Agresti2020, Gallego2013,Brandao2016,liu2018device,nieto2018device} and our algorithm is able to find them, our aim here is to maximize the certified randomness directly from the data, instead of an a-priori known Bell inequality~\cite{nieto2014using}. Following \cite{pironio2013security} we have approached this problem by constraining the guessing probability of the adversary (Eve) directly on the observed behaviour. Denoting as $e$ the outcome associated with Eve, randomness can be quantified via the guessing probability~\cite{note3} given by
\begin{equation}
    p_{\mathrm{guess}}(x,y) = \max_{\{p \in \mathcal{Q}\}} \sum_{a, b} p(a, b,e=(a,b) | x,y), 
\end{equation}
where $p(a,b,e|x,y) \in \mathcal{Q}$ is the unknown global behavior including Eve, constrained to belong to the set of quantum correlations $\mathcal{Q}$.
The distribution $p_{\mathrm{guess}}$ represents the amount of knowledge that she can extract on the outcomes of Alice and Bob, and it must be compatible with the observed distribution $p^*(a,b|x,y) = \Tr(A_{a|x} \otimes B_{b|y} \; \rho)$.
Combining the stochastic Nelder-Mead algorithm with the Navascues-Pironio-Acin hierarchy \cite{navascues2007bounding} we are able to optimize the upper bound on $p_\mathrm{guess}$ over unknown measurements and states constrained on being compatible 
with $p^*$, i.e. we are solving the following maximization problem:
\begin{equation}
\begin{split}
    \max~ & p_\mathrm{guess}\\
    \textrm{s.t.} ~ & p(a,b,e|x,y) \in \mathcal{Q}_k\\
    & \sum_{e} p(a,b,e|x,y) = p^*(a,b|x,y)    
    \end{split}
\label{sdp_prob2}
\end{equation}
where $\mathcal{Q}_k$ represent the relaxation of quantum set at the $k$ order of the NPA hierarchy~\cite{navascues2007bounding}.
In Fig.~\ref{fig:SIrandomness} we show, for states of the form \eqref{eq:stategen} with different values of the parameter $\gamma$, how our algorithm is able, in less than 500 iterations, to approach the optimal value of the $p_{guess}$  for pure states~\cite{nieto2014using}.
Considering experimental density matrices, we are able to approach the higher value of $p_{guess} \approx 0.7$ as is expected due to the noise of experimentally reconstructed states.

\begin{figure}[t!]
    \includegraphics[width=.9\columnwidth]{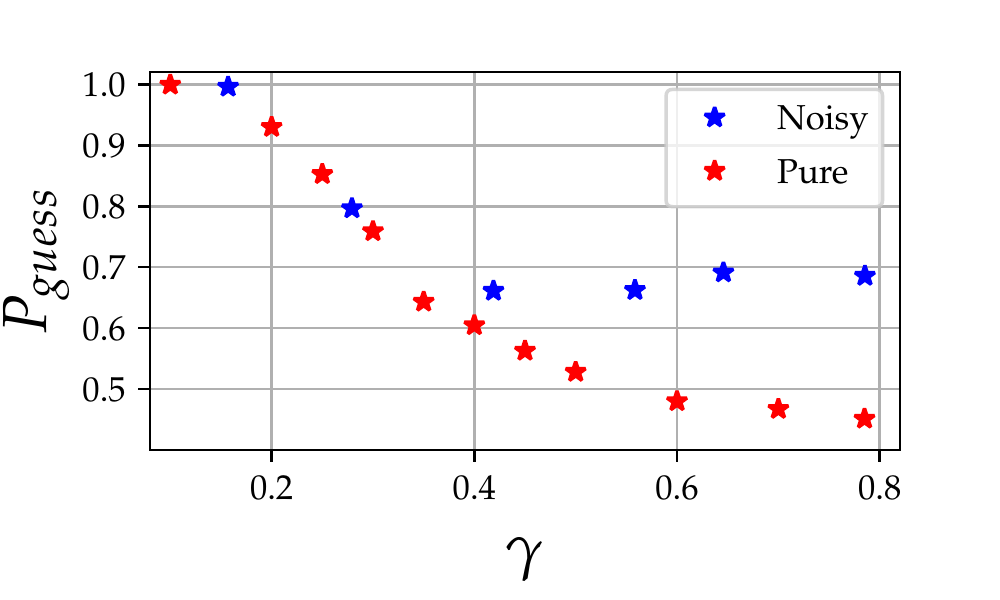}
\caption{\textbf{Simulated Randomness Certification.} Optimal $p_{guess}$, obtained as average over 50 runs of the optimal value reached by our algorithm after 500 iterations, as a function of the parameter $\gamma$ for pure states in the form \eqref{eq:stategen} (red stars) and for experimentally reconstructed noisy density matrices (blue stars). The noisy points can certify less randomness than the ideal ones. The trend of $p_{guess}$ in Eq. \eqref{sdp_prob2} as a function of the iterations. 
}
\label{fig:randomness}
\end{figure}

The simulation results are reported in Fig.~\ref{fig:randomness} and clearly show that our approach is able to directly maximize the randomness extraction in a fully black-box approach.

\section{Discussion}
The violation of a Bell inequality is often pictured as the paradigmatic example of a device-independent task: from the observed statistic alone, without knowing the internal working of devices, one can conclude the nonclassical nature of it. To obtain such violation, however, one often has to rely on a precise description of the devices. Here we propose a practical solution to this problem, exploiting an adaptive automated algorithm able to maximize the Bell inequality violation in a fully black-box setting. Employing both simulation and actual experiments, we demonstrated the protocol by optimizing the violation of many Bell inequalities for different unknown photonic bipartite states and measurement responses (see also \cite{supp}). Nicely, the optimum values are achieved after a few hundred iterations.

Our approach can also be applied to quantum networks of growing size and complexity that have started to be experimentally and theoretically explored in the recent years \cite{branciard2012bilocal,chaves2015information,fritz_2012, fritz2016beyond, renou_2019, gisin2019entanglement,chaves2021causal,carvacho2017experimental,saunders2017experimental,sun2019experimental,poderini2020experimental,tavakoli2021bell,agresti2021experimental}. Finding Bell inequalities in these cases can be very difficult, but heuristic machine learning based approaches have been proposed to find and quantify nonclassicality \cite{krivachy2020neural,canabarro2019machine,krivachy2020fast} opening the possibility of applying our framework also in such scenarios.  
Moreover, practical experimental quantum information tasks, where multiple parameters are tuned to optimize a desired cost function, can also benefit of our approach. 
To demonstrate that, we have applied our framework for a paradigmatic application of Bell's theorem, showing that even with no knowledge of states nor devices, one can directly (i.e., without the need of Bell inequalities) maximize the ammount of certified randomness.

\begin{acknowledgments}
\textit{Acknowledgments--}
This work was supported by The John Templeton Foundation via the grant Q-CAUSAL No 61084 and via The Quantum Information Structure of Spacetime (QISS) Project (qiss.fr) (the opinions expressed in this publication are those of the author(s) and do not necessarily reflect the views of the John Templeton Foundation) Grant Agreement No. 61466, and by the ERC Advanced Grant QU-BOSS (Grant agreement no. 884676). RC acknowledges the Serrapilheira Institute (Grant No.  Serra-1708-15763), CNPq via the INCT-IQ and Grants No.307172/2017-1 and No.406574/2018-9 and Brazilian agencies MCTIC and MEC.
\end{acknowledgments}

\clearpage

\appendix

\section{Details of the Stochastic Nelder Mead algorithm}

The Stochastic Nelder Mead algorithm is composed of three main steps, continuously repeated during the optimization process:  

\begin{enumerate}

 \item The Bell inequality parameter $S$ is calculated for all the points $\bm{t}$ of the simplex $\Omega_{k-1}$ and the point with the maximum value is removed from the simplex, if $dim(\Omega_{k-1})>n$. Then the values $S(\bm{t})$ are sorted and three elements are individuated: $\bm{t}^{max}$, $\bm{t}^{2nd\,max}$ and $\bm{t}^{min}$ that are the points for which the CHSH parameter $S(\cdot)$ assumes the maximum, the second maximum  and the minimum values, respectively.

\item The barycenter $\bm{t}^{bar}$ of the points in $\Omega_{k-1}  \setminus \{\bm{t}^{max}\}$ is calculated and a new point $\bm{t}^{ref}$ is generated by the following reflection rule: 
\begin{centering}
$\bm{t}^{ref}=(1+\delta)\bm{t}^{bar}-\delta \, \bm{t}^{max} $
\end{centering}

where $\delta >0$ is the reflection coefficient.

\item 
\begin{enumerate}

\item  If $S(\bm{t}^{min}) \le S(\bm{t}^{ref})< S(\bm{t}^{2nd max})$, impose $\Omega_{k}=\Omega_{k-1} \cup \{\bm{t}^{ref} \}$.

\item If $ S(\bm{t}^{ref})< S(\bm{t}^{min})$, generate the expansion $\bm{t}^{exp}=\gamma \bm{t}^{ref}+ (1-\gamma) \, \bm{t}^{bar} $
where $\gamma >1$ is the expansion coefficient
 and if $ S(\bm{t}^{exp})< S(\bm{t}^{ref})$ impose $\Omega_{k}=\Omega_{k-1} \cup \{\bm{t}^{exp} \}$, otherwise $\Omega_{k}=\Omega_{k-1} \cup \{\bm{t}^{ref} \}$.
 
\item If $ S(\bm{t}^{ref}) \ge S(\bm{t}^{2nd max})$, then
\begin{enumerate}
\item If $S(\bm{t}^{2ndmax}) \leq S(\bm{t}^{ref}) < S(\bm{t}^{max})$ there will be an \textit{external} contraction given by
\begin{align*}
\bm{t}^{cont} = \beta\bm{t}^{ref} + (1 - \beta)\bm{t}^{cent} \\
\qquad 0 \leq \beta \leq 1.
\end{align*}
If $S(\bm{t}^{cont}) \leq S(\bm{t}^{ref})$ the contraction is accepted.\\

    \item If $S(\bm{t}^{ref}) \geq S(\bm{t}^{max})$ there will be an \textit{internal} contraction given by
    \begin{align*}
    \bm{t}^{cont} = \beta\bm{t}^{max} + (1 - \beta)\bm{t}^{cent} \\\qquad 0 \leq \beta \leq 1.
    \end{align*}
If $S(\bm{t}^{cont}) \leq S(\bm{t}^{max})$ the contraction is accepted.\\
\end{enumerate}
\end{enumerate}
\end{enumerate}

If the contraction is accepted then $\Omega_{k+1} = \Omega_{k} \cup \{\bm{t}^{cont}\}$, otherwise an \textit{Adaptive Random Search} (ARS) is exploited as described in the main text.

\begin{figure}[ht]
\includegraphics[width=\columnwidth]{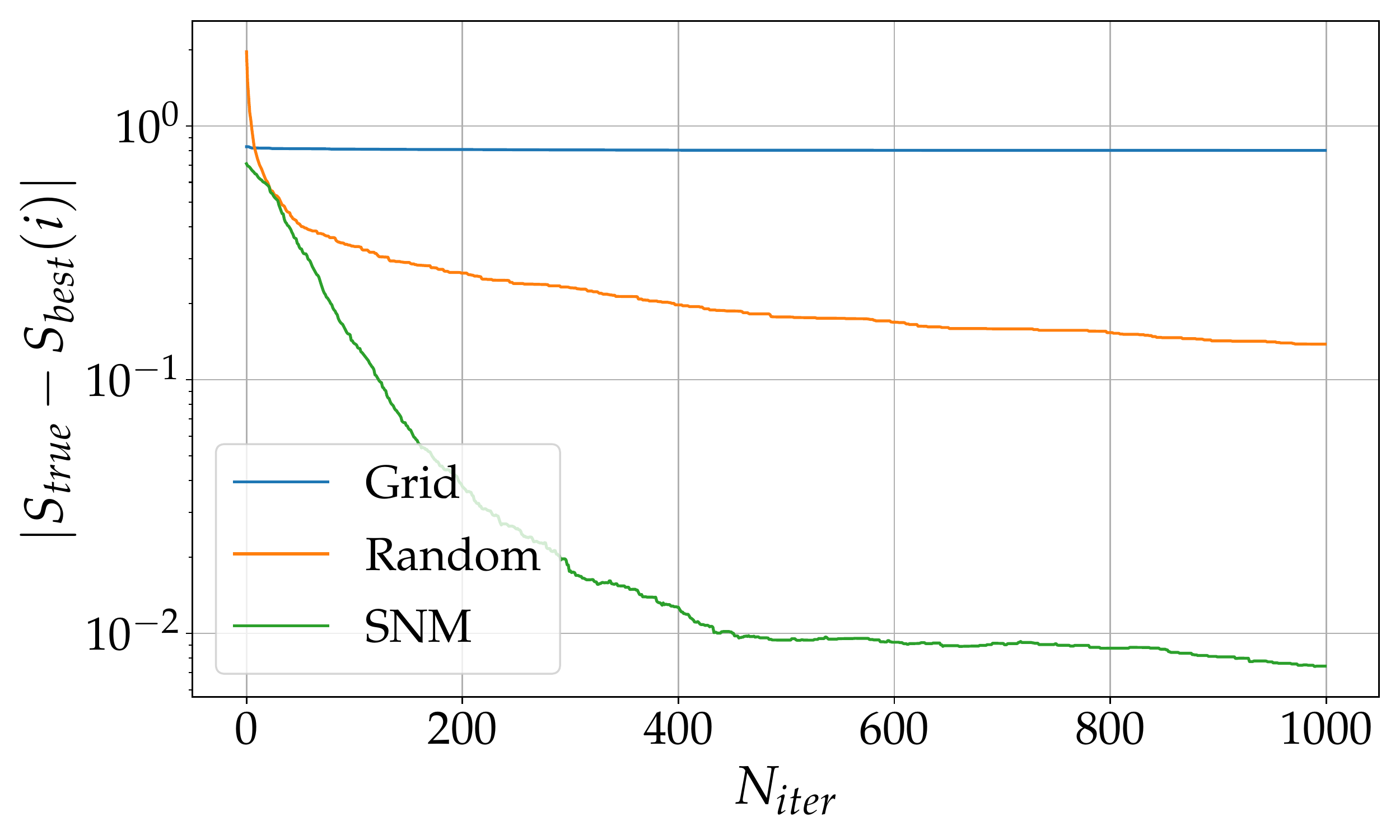}
\caption{\textbf{Comparison between different optimization procedures (simulation).} The plot considers the $8$-dimensional parameter space of the CHSH scenario.  The grid optimization (blue curve) is a brute force search on an equally spaced grid in this space. The plot shows the average of $100$ runs where the starting point of the grid was chosen at random. The random search method (orange curve) instead performs sampling on random points trying to find the optimum. Also in this case the average of $100$ runs is shown. As can be seen the Stochastic Nelder Mead algorithm (green curve) performs much better.
All curves were generated by simulating noisy samples following Poissonian statistics corresponding to $N=10^4$ events.}
\label{fig:gridran}
\end{figure}

\begin{figure*}[ht]
\includegraphics[width=\textwidth]{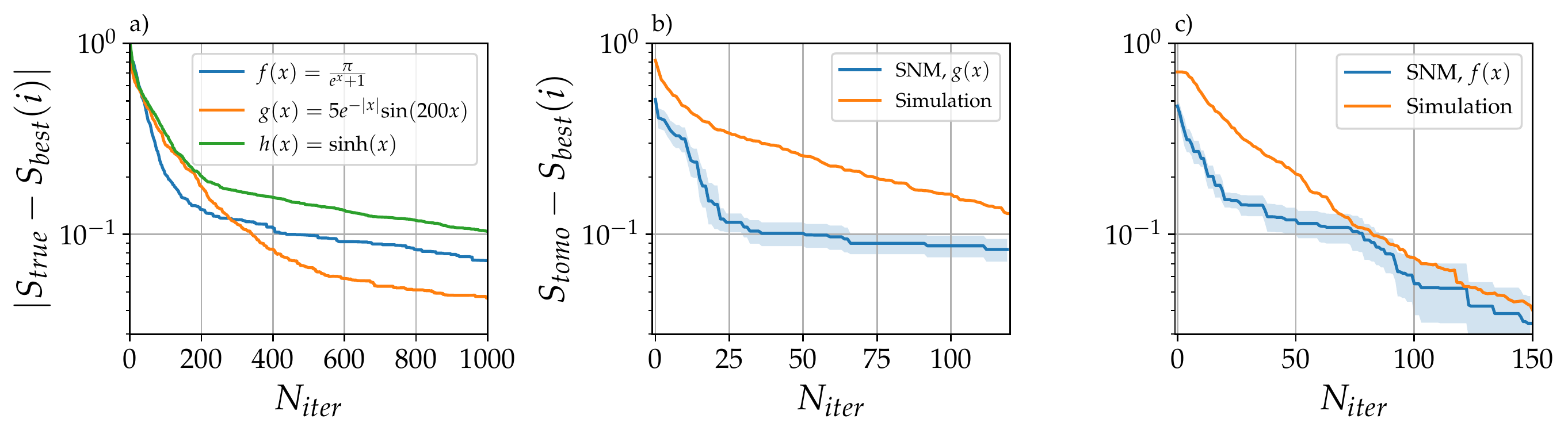}
\caption{\textbf{Optimization with different measurement responses (simulation and experiment).}
\textbf{a)} The difference $|S_{true}-S_{best}|$, averaged over 100 runs, is plotted as function of the algorithm's iteration. Here the optimization of the CHSH inequality is performed over pure singlet states using non linear maps to the $8$ total projector parameters, accounting for Poissonian fluctuations with $N=10^4$. 
\textbf{b-c)} Mean difference $S_{tomo} - S_{best}$ as a function of the algorithm's iteration considering the $\textrm{Osc}(\alpha)$ and $\textrm{Logi}(\alpha)$ non linear maps as measurement response in both Alice's and Bob's station. In both cases the QWP position has been fixed on its optical axis; the algorithm is controlling only the HWP. For both figures b) and c) the mean rate of events per measurement is $\approx 5\cdot10^4$ and $9$ experimental runs have been collected.
 The orange lines indicate the average simulations (repeated 100 times) for  a Werner state with $p=0.90$, employing  $N=5 \cdot 10^4$ Poissonian events with added Gaussian noise ($\sigma=0.05$)  to simulate the real experimental conditions. In figure b) we have obtained $S_{tomo}=2.574 \pm 0.015$ and our average over the optimization runs is $S_{best}=2.491 \pm 0.016$ with a variance of $0.0023$, while in figure c) we have $S_{tomo}=2.469 \pm 0.017$ and $S_{best}=2.435 \pm 0.026$ with a variance of $0.006$.
}
\label{fig:bbfunc}
\end{figure*}

The standard control parameters used in our algorithm are: $\{\delta, \eta, \gamma\} = {1, 0.5, 2}$  \cite{nelder1965simplex} and $\epsilon$ equal to $\approx 10\%$ of the value of the parameters variation used to perform the ARS global search process.

\section{Comparison between Stochastic Nelder Mead algorithm and non-adaptive algorithms}
In this section we compare our approach to  two basic non-adaptive procedures for gradient-free optimization in high dimensional parameter space.

The first is a brute force approach where the function
is optimized on a equally spaced grid in the parameter space. The exponential increase in the number of points in the grid given by $n^d$ -- where d is the number of optimization parameter
(for instance, $d=8$ for the CHSH case) and $n$ is the number of samples for a single parameter-- severely limits the applicability of this method for high parameter space. As a matter of fact,  usually this approach is used in combination with others local optimization methods. To illustrate the method, here we employ $n=3$ which gives $6561$ grid points to sample.

Among the simplest and effective non-adaptive approaches to multiparameter black-box optimization is to perform a random search in the parameter space.
Here, at each step, sets of parameters are selected uniformly at random in the allowed set and independently in respect to previous steps.

Employing such approaches and comparing them to our optimization procedure shows a definite advantage of the latter. 
In particular the convergence to the optimum is notably faster, a feature of fundamental importance for real applications where the number of samples one can afford is severely limited.

For comparison we performed both the grid and random optimization simulating noisy measurements with Poissonian statistics corresponding to $N=10^4$ events.
The results are shown in Fig. \ref{fig:gridran}, considering the average of $100$ optimization procedures, where in the grid case we randomized the starting point of the grid in the parameter space.

\begin{figure}[ht!]
 \includegraphics[width=.9\columnwidth]{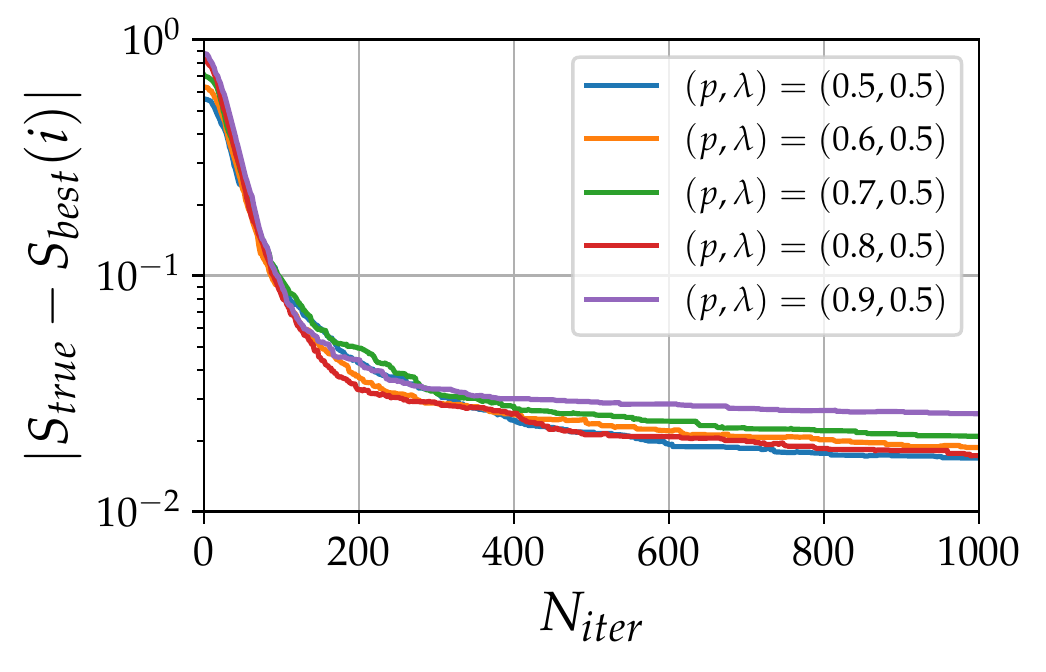}
\caption{ \textbf{Simulations for noisy states.} Simulated scenario with different mixed states, parametrized by $p$ and $\lambda$, with a number of events $N=10^4$.
The difference $|S_{true}-S_{best}|$ is plotted as function of the algorithm's iteration. All the curves are obtained by averaging over 100 simulation runs.
}
\label{fig:noisy}
\end{figure}

\section{CHSH inequality optimization with different measurement responses and noisy states}

In the our photonic implementation with qubits each projective measurement is described by two continuous angles (from the half and quarter wave plates). In the black-box scenario, however, we cannot assume how the actual measurement operator depends on such inputs.
To illustrate the role the choice of the response functions might have in the optimization we considered the following three response functions:
\begin{eqnarray*}
f(\alpha) =
\left \{
\begin{array}{rl}
 5e^{-|\alpha|}\sin(200\alpha) & \equiv \text{Osc}(\alpha), \\
\frac{\pi}{e^\alpha + 1} & \equiv \text{Logi}(\alpha),\\
\sinh(\alpha) .
\end{array}
\right.
\end{eqnarray*}

As can be seen in Fig. \ref{fig:bbfunc}a) all such response functions reach a reasonable accuracy in our simulations. As shown in Fig. \ref{fig:bbfunc}b-c), we compare the numerical simulations with an actual experiment, by choosing the oscillating and logistic response functions. The experimental minimization is done by fixing the QWP positions and optimizing on the 4 measurement parameters associated with the HWP positions. 

Furthermore, we simulate the optimization of  CHSH inequality for different states of the form: $ \rho= p |\psi^{\gamma} \rangle \langle \psi^{\gamma}|+ 
    + (1-p) \left(
    \lambda \frac{|\psi^{-}\rangle \langle \psi^{-}|+|\psi^{+}\rangle \langle \psi^{+}|}{2}+(1-\lambda) \frac{ \mathbb{I}}{4}
    \right)$, with different values of noise parameters $p$ and $\lambda$. The results are shown in Fig. \ref{fig:noisy} and demonstrate how the algorithm can optimize all the tested states with similar performances.


\begin{thebibliography}{75}
\bibitem [{not({\natexlab{a}})}]{note1}%
  \BibitemOpen
  \href@noop {} {} \bibinfo {note} {A simplex is the
  $n$-dimensional generalization of the triangle. Given $n-1$ measurement
  parameters, a simplex contains $n$ vertices that correspond to the parameters
  points in the $n-1$ dimensional space.}\BibitemShut {Stop}%
\bibitem [{not({\natexlab{b}})}]{note2}%
  \BibitemOpen
  \href@noop {} {} \bibinfo {note} {The goal of the ARS is
  to sample the space of measurement parameters using a local or global search,
  in order to find a point able to improve the simplex, when the contraction
  step does not provide a satisfactory point. The process is randomly driven:
  with a probability $P_s > 0$ the Bell parameter $S$ is evaluated in one point
  uniformly chosen in the parameter space (global search), while with
  probability $1-P_s > 0$ a point of the simplex is randomly chosen and a point
  $\bm{t}_i$ is uniformly extracted from the hypersphere with center $\bm{t}_i$
  and radius $\epsilon$ (local search). If $S(\bm{t}^{ARS}) \leq
  S(\bm{t}^{max})$ the search stops, otherwise the ARS is
  repeated.}\BibitemShut {Stop}%
\bibitem [{not({\natexlab{c}})}]{note4}%
  \BibitemOpen
  \href@noop {} {} ({\natexlab{c}}),\ \bibinfo {note} {this value becomes 1.6
  bits if we employ the Von Neumann entropy instead of the min-entropy to
  characterize the generated randomness.}\BibitemShut {Stop}%
\bibitem [{not({\natexlab{d}})}]{note3}%
  \BibitemOpen
  \href@noop {} {} ({\natexlab{d}}),\ \bibinfo {note} {the guessing probability
  is directly connected to the conditional min-entropy
  $H_{min}=-log_2(p_\mathrm{guess})$ representing the bits of randomness
  certifiable by Alice and Bob.}\BibitemShut {Stop}%
\end{thebibliography}
\end{document}